\documentclass[%
 reprint,
nofootinbib,
 amsmath,amssymb,
 aps,
pra,
]{revtex4-1}

\usepackage{graphicx}% Include figure files
\usepackage{dcolumn}% Align table columns on decimal point
\usepackage{bm}% bold math
\usepackage{epsfig}
\usepackage{subfigure}
\usepackage{color}

\begin{document}

\preprint{APS/123-QED}

\title{\boldmath Searches for isospin-violating transitions $\chi_{c0,2} \rightarrow \pi^{0} \eta_{c}$}

% \author{Olga Bondarenko}
% %\email{o.bondarenko@rug.nl}
% \affiliation{KVI}

\author{
\begin{small}
\begin{center}
M.~Ablikim$^{1}$, M.~N.~Achasov$^{8,a}$, X.~C.~Ai$^{1}$, O.~Albayrak$^{4}$, M.~Albrecht$^{3}$, D.~J.~Ambrose$^{43}$, 
A.~Amoroso$^{47A,47C}$, F.~F.~An$^{1}$, Q.~An$^{44}$, J.~Z.~Bai$^{1}$
% ...
\\
\vspace{0.2cm}
(BESIII Collaboration)\\
\vspace{0.2cm} {\it
$^{1}$ Institute of High Energy Physics, Beijing 100049, People's Republic of China\\
% ...
$^{3}$ Bochum Ruhr-University, D-44780 Bochum, Germany\\
$^{4}$ Carnegie Mellon University, Pittsburgh, Pennsylvania 15213, USA\\
% ...
$^{8}$ G.I. Budker Institute of Nuclear Physics SB RAS (BINP), Novosibirsk 630090, Russia\\
% ...
$^{43}$ University of Rochester, Rochester, New York 14627, USA\\
$^{44}$ University of Science and Technology of China, Hefei 230026, People's Republic of China\\
% ...
$^{47}$ (A)University of Turin, I-10125, Turin, Italy; (B)University of Eastern Piedmont, I-15121, Alessandria, Italy; (C)INFN, I-10125, Turin, Italy\\
% ...
$^{a}$ Also at the Novosibirsk State University, Novosibirsk, 630090, Russia\\
% ...
}\end{center}
\vspace{0.4cm}
\end{small}
}

\collaboration{BESIII Collaboration}

\date{\today}

\begin{abstract}
We present the first upper-limit measurement of the branching fractions of the isospin-violating transitions $\chi_{c0,2} \rightarrow \pi^{0} \eta_{c}$. 
The measurements are performed using $106\times 10^{6}$ $\psi(3686)$ events accumulated with the BESIII detector at the BEPCII $e^{+}e^{-}$ 
collider at a center-of-mass energy corresponding to the $\psi(3686)$ mass. 
We obtained upper limits on the branching fractions at a 90\% confidence level of 
$B(\chi_{c0} \rightarrow \pi^{0} \eta_{c}) < 1.6 \times 10^{-3}$ and 
$B(\chi_{c2} \rightarrow \pi^{0} \eta_{c}) < 3.2 \times 10^{-3}$.
%, with branching fractions of $\psi(3686) \rightarrow \gamma \chi_{c0,2}$ and $\eta_{c} \rightarrow K^{0}_{S}K^{\pm}\pi^{\mp}$ 
%taken from the Particle Data Group.

\end{abstract}

\maketitle

\section{Introduction}

Isospin is known to be a good symmetry in the hadronic decays of charmonium states. %below the $D\bar{D}$-production threshold. 
The decay rates of isospin-symmetry breaking modes are
in general found to be very small. For example, the branching fraction~($B$) of the
measured isospin-violating transition $\psi(3686)\rightarrow \pi^{0} J/\psi$ was found
to be only $(1.26 \pm 0.02 (stat.) \pm 0.03 (syst.)) \times 10^{-3}$~\cite{pi0jpsi2012}, whereas for other hadronic transitions such
as $\psi(3686)\rightarrow \pi^{+}\pi^{-} J/\psi$, the branching fraction is $(34.45 \pm 0.30)\times 10^{-2}$~\cite{PDG2014} and thus significantly stronger.

Although isospin breaking is found to be very small for the conventional charmonium states,
the mysterious $X(3872)$ resonance above the $D\bar{D}$ threshold decays strongly 
via the transition $X(3872)\rightarrow \pi^{+}\pi^{-} J/\psi$, where the
invariant-mass spectrum of the $\pi^{+}\pi^{-}$ pair shows a clear $\rho$ signature~\cite{X3872Belle2003isospin,BELLEX3872,CDFX3872,CMSJHEP04}
and, hence, is compatible with an isospin-violating decay.
A possible interpretation is that the $X(3872)$ is a molecular state composed of a bound $D^{*0}$-$\bar{D}^0$
meson pair (\cite{Close2004119, Tornqvist2004209, Voloshin2004316, Swanson2004197}). 
Such an explanation is particularly popular, since the mass of the
$X(3872)$ is close to the sum of the ${\bar D}^0$ and $D^{*0}$ masses, pointing to
a state that could be weakly bound by the exchange of a color-neutral meson, similar to the deuteron. 
Moreover, in such a scenario, the strong isospin-breaking decay rate of the $X(3872)$ might be explained by the large mass gap between
the $D^{*0}-\bar{D}^0$ and the $D^{*+}-D^-$ ($D^+-D^{*-}$) thresholds~\cite{Li2012}. 
%A better understanding of the isospin-breaking mechanism in a controlled system,
%such as charmonium, could be crucial to shed light on the nature of the $X(3872)$.
A better understanding of the isospin-breaking mechanism in a complementary and well-established charmonium system below 
the open-charm threshold could be crucial to shed light on the nature of the $X(3872)$.

%At leading order of the QCD multipole expansion~\cite{Ioffe1980}
% \begin{equation}
% \begin{split}
%  R_{\pi^{0}/\eta} = 
% \frac{B(\psi(3686) \rightarrow \pi^{0}J/\psi)}{B(\psi(3686) \rightarrow \eta J/\psi)} = \\
% 3\left(\frac{m_{d} - m_{u}}{m_{d} + m_{u}}\right)^{2} \frac{F^{2}_{\pi}}{F^{2}_{\eta}} \frac{M^{4}_{\pi}}{M^{4}_{\eta}} \left | \frac{\overrightarrow{q}_{\pi}}{\overrightarrow{q}_{\eta}} \right |^{3},
% \label{eq:memo1}
% \end{split}
% \end{equation}
% where $\overrightarrow{q}_{\pi(\eta)}$ is the pion(eta) momentum in the rest frame of $\psi(3686)$, 
% $F_{\pi(\eta)}$ and $M_{\pi(\eta)}$ are the decay constant and mass of the pion (eta), respectively. 
%Using Eq.~\ref{eq:memo1} and the CLEO measurement of the decay-width ratio~\cite{CLEOpi0Jpsi}, 

On the quark level, the isospin-symmetry is broken due to the electromagnetic interaction 
and due to differences in the up- and down-quark masses ($m_{u}$ and $m_{d}$). 
It is, therefore, believed that isospin-breaking decays can be used to access 
the up- and down-quark mass differences once the electromagnetic effect is either well understood or found to be negligible. 
An example observable that has been proposed to obtain the quark mass ratio, $m_{u}/m_{d}$, is a measurement of the ratio between 
the branching fractions of the transitions $\psi(3686) \rightarrow \pi^{0}J/\psi$ and $\psi(3686) \rightarrow \eta J/\psi$. 
Based on a leading-order QCD multipole expansion~\cite{Ioffe1980} and the BESIII measurement of this ratio~\cite{pi0jpsi2012}, 
the up-down quark mass ratio is extracted to be $m_{u}/m_{d} = 0.407 \pm 0.006$. 
This result is smaller than the result, $m_{u}/m_{d} = 0.56$, obtained using the Goldstone boson masses 
from a leading-order chiral-perturbation theory~\cite{Weinberg1977}. It is important to understand such a 
large discrepancy between the values of $m_{u}/m_{d}$ obtained on the basis of different theoretical conjectures. 

The most promising developments in this field are based upon an effective-field theoretical approach.
A nonrelativistic effective-field theoretical (NREFT) study by the J\"{u}lich and 
IHEP groups suggests that intermediate (virtual) charmed-meson loops are the dominant source for the isospin 
breaking in the transition $\psi(3686)\rightarrow \pi^{0}J/\psi$~\cite{Hanhart2009, ErratumHanhart}. 
According to the proposed theory, the contribution of charmed-meson loops to the amplitude of the process 
is enhanced by a factor of $(\upsilon/c)^{-1} \sim 2$, where $\upsilon$ is the heavy-meson velocity in the loops.
%For the discussed transition, $(\upsilon/ c) \backsimeq 0.53$.
Detailed studies of different isospin-violating transitions in charmonium below the $D\bar{D}$ threshold and the effect of 
virtual charmed-meson loops on the widths of the transitions are described in Ref.~\cite{Hanhart2010}.  

The NREFT calculations described above are based on a first estimate, exploiting diagrams involving the lowest-lying pseudoscalar 
and vector charmed mesons following heavy-quark symmetry and chiral symmetry.
Although these theoretical calculations give qualitative insights in the isospin-breaking mechanisms in charmonium decays, 
the authors in Ref.~\cite{Hanhart2010} state that only with a further developed effective-field theory that includes
Goldstone bosons, charmonia, and charmed mesons as the degrees of freedom, it would be possible in the future to extract 
the light-quark masses from quarkonia decays. Currently, for such a theory, quantitative predictions of individual 
branching fractions of isospin-forbidden decays of charmonium are difficult, 
because information on the coupling constants $f_{\psi D\bar{D}}$ between different charmonium states and $D\bar{D}$-mesons is limited. 
The theory requires constraints from experimental data, in particular from measurements of decay rates of other isospin-violating
transitions in charmonium~\cite{Hanhart2010}. 
%An input from Lattice QCD, in particular a calculation of $f_{\psi D\bar{D}}$, is also needed. 
%An advanced effective field theory will hopefully lead to a rigorous tool to reveal the nature of the XYZ states, such as the $X(3872)$.

In this paper, we present an experimental study of the isospin-suppressed transition of the charmonium P-wave states $\chi_{c0,2}$ 
to the ground state $\eta_c$ via the emission of the $\pi^{0}$. The $\chi_{c0,2}$ states are obtained 
via electromagnetic transitions, $\psi(3686)\rightarrow \gamma \chi_{c0,2}$, whereby the $\psi(3686)$ resonance is directly populated via the $e^+e^-$ annihilation process. 
The transition $\chi_{c1} \rightarrow \pi^{0} \eta_{c}$ is not considered in this analysis since it violates conservation of parity and 
angular momentum.
According to Ref.~\cite{Hanhart2010}, the dimensionless suppression factor for the loops in $\chi_{c0} \rightarrow \pi^{0} \eta_{c}$ is 0.2. 
This factor is smaller than in the process $\psi(3686) \rightarrow \pi^{0} J/\psi$, however, through the interference 
with the tree-level amplitude, meson loops may give a significant contribution and cannot be neglected.    
%The upper limits on the branching fractions $B(\chi_{c0,2} \rightarrow \pi^{0} \eta_{c})$ are obtained for the first time.

\section{The BESIII experiment and data set}

The analysis is based on the $\psi(3686)$ data sample accumulated by the BESIII detector in 2009. 
The total number of $\psi(3686)$ events is $(106.41 \pm 0.86)\times 10^{6} $~\cite{NumPsip}, corresponding to an integrated luminosity of 156.4~pb$^{-1}$. 
In addition, 42.6 pb$^{-1}$ data collected at a center-of-mass energy of 3.65~GeV, are used to estimate the background from nonresonant processes.

The BEijing Spectrometer III (BESIII), described in detail in Ref.~\cite{BESIIINIM}, 
is a detector for $\tau$-charm studies running at the Beijing Electron-Positron Collider (BEPCII). 
BEPCII is a double-ring $e^{+}e^{-}$ collider with a designed peak luminosity of $10^{33}$~cm$^{-2}$s$^{-1}$ at a beam current of 0.93~A. 
The cylindrical core of the BESIII detector consists of a main drift chamber (MDC), 
a plastic scintillator time-of-flight system (TOF), and an electromagnetic calorimeter (EMC), 
which are enclosed in a superconducting solenoidal magnet providing a 1~T magnetic field. 
The solenoid is supported by an octagonal flux-return yoke with resistive-plate chambers forming a muon counter system. 
The MDC is a small-cell, helium-based (40\% He, 60\% C$_{3}$H$_{8}$) subdetector consisting of 43 layers and providing 
an average single-hit resolution of 135~$\mu$m, and a charged-particle momentum resolution of 0.5\% at 1~GeV/$c$. 
The EMC subdetector consists of 6240 CsI(Tl) crystals in a cylindrical structure (barrel) and two end caps. 
For 1~GeV photons, the energy resolution is 2.5\% (5\%) and the position resolution is 6~mm (9~mm) for the barrel (end caps).
The TOF system consists of 5~cm thick scintillators, with 176 detectors of 2.4~m length 
in two layers in the barrel and 96 fan-shaped detectors in the end caps. 
The barrel (end cap) time resolution of 80~ps (110~ps) provides $2\sigma$ $K/\pi$ separation for momenta up to 1~GeV. 

To optimize the event selection, to estimate background contributions, and to evaluate the
detection efficiencies, Monte Carlo (MC) simulated samples are obtained exploiting 
a realistic model 
of the detector. For this, the GEANT4-based simulation software BOOST~\cite{BOOST} is used which includes the geometry and material 
description of the BESIII spectrometer, and the detector response.  
A MC sample based on 106~M inclusive $\psi(3686)$ decays is used to study the background. 
This inclusive sample is generated with KKMC~\cite{KKMC} plus EvtGen~\cite{EvtGenPaper,BesEvtGen} 
and the known branching ratios are taken from the Particle Data Group (PDG)~\cite{PDG2014}, while
the unknown ratios are generated according to the Lundcharm model~\cite{Lundcharm}. 
The decay modes $\chi_{c0,2} \rightarrow \pi^{0} \eta_{c}$ are not present in the inclusive MC simulation.
Signal MC samples are generated to determine the detection efficiency and to model the signal shape. 
In the MC simulations for the processes presented here, the $\psi(3686)\rightarrow\gamma\chi_{cJ}$ decay 
is assumed to be a pure E1 transition, and the polar angle, $\theta$, follows a distribution of the form 
$1 +\alpha\cos^{2}\theta$, with $\alpha= 1$ and 1/13 for $J= 0$ and 2, respectively ~\cite{AngChi1, AngChi2}.  
The $\chi_{c0,2}\rightarrow\pi^{0}\eta_{c}$ and $\eta_{c}\rightarrow K^{0}_{S}K^{\pm}\pi^{\mp}$ 
decays are assumed to be pure phase-space decays. 

%%%%%%%%%%%%%%
% Label of figures match caption?
%%%%%%%%%%%%%%

\begin{figure*}[ht]
 \centering
 \includegraphics[scale=0.33]{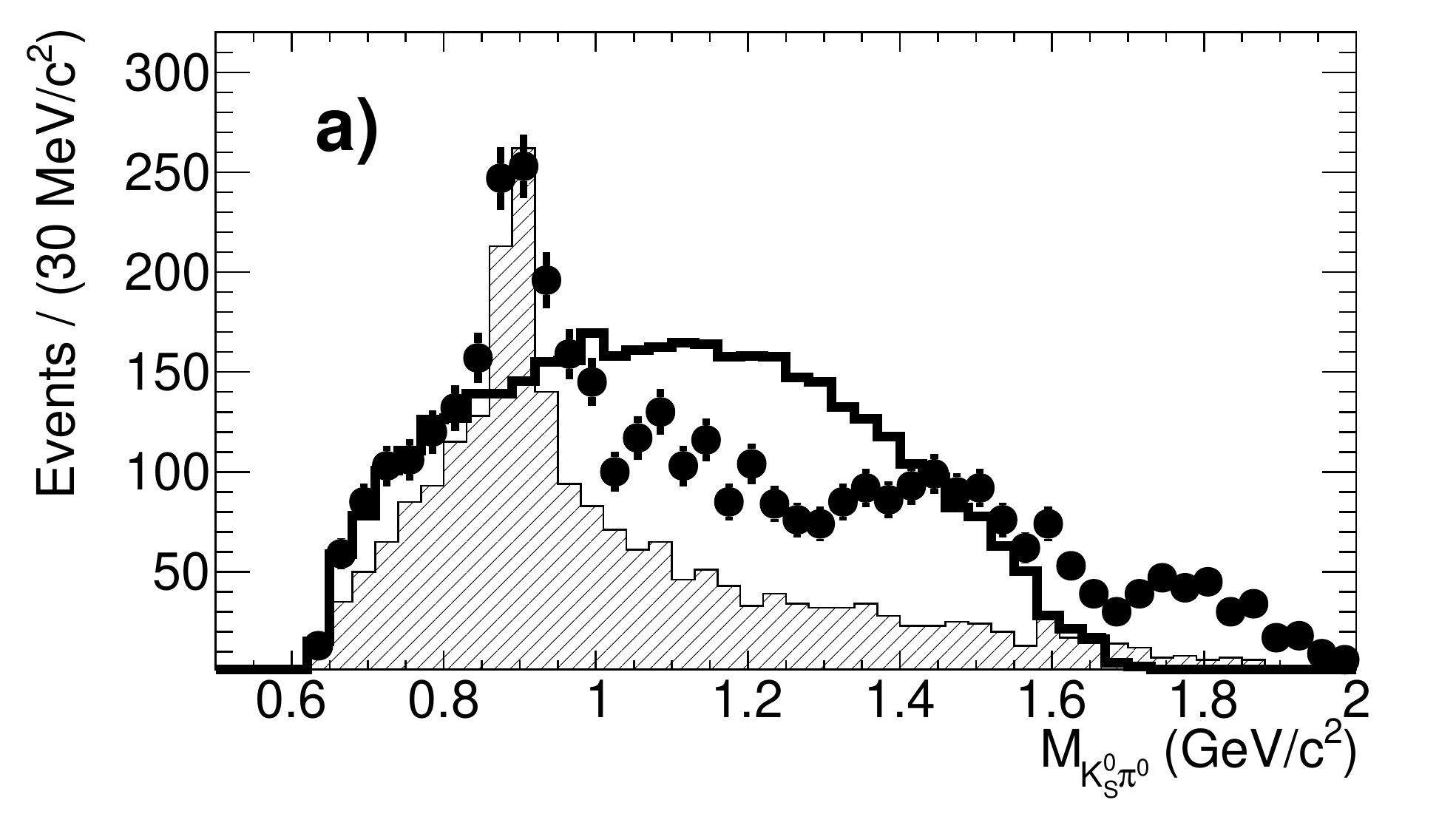}
 \includegraphics[scale=0.33]{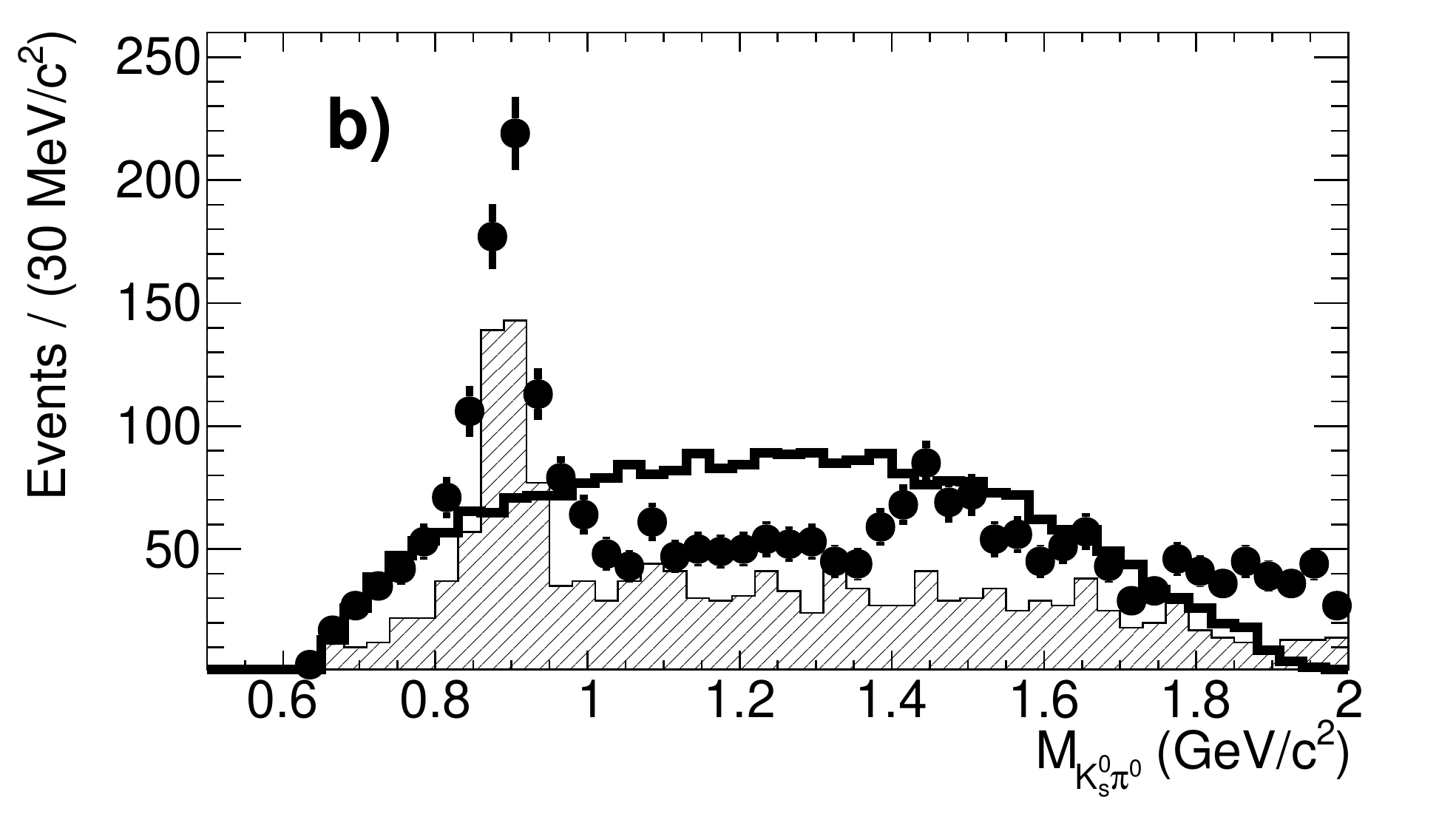}\\
 \includegraphics[scale=0.33]{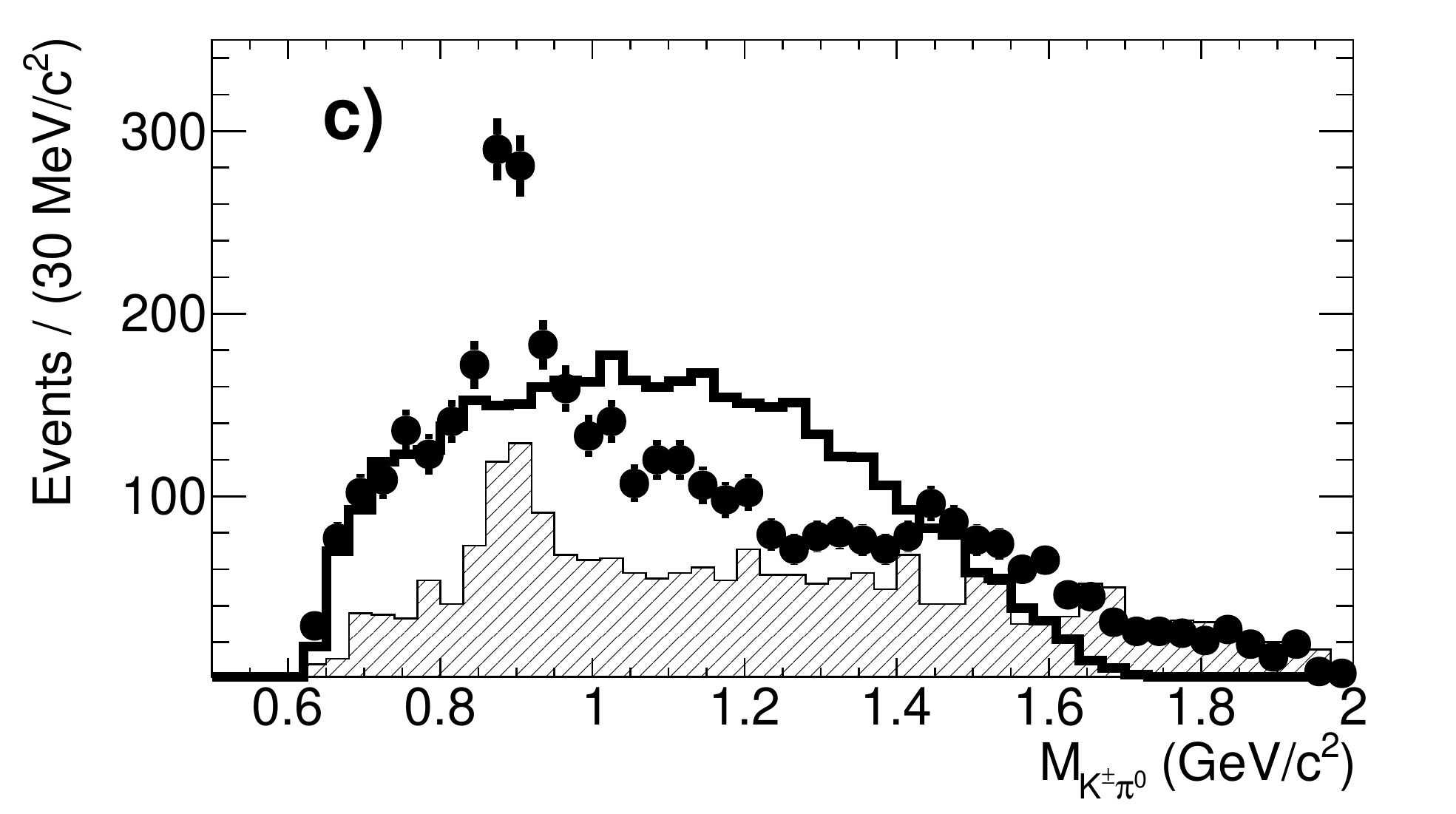}
 \includegraphics[scale=0.33]{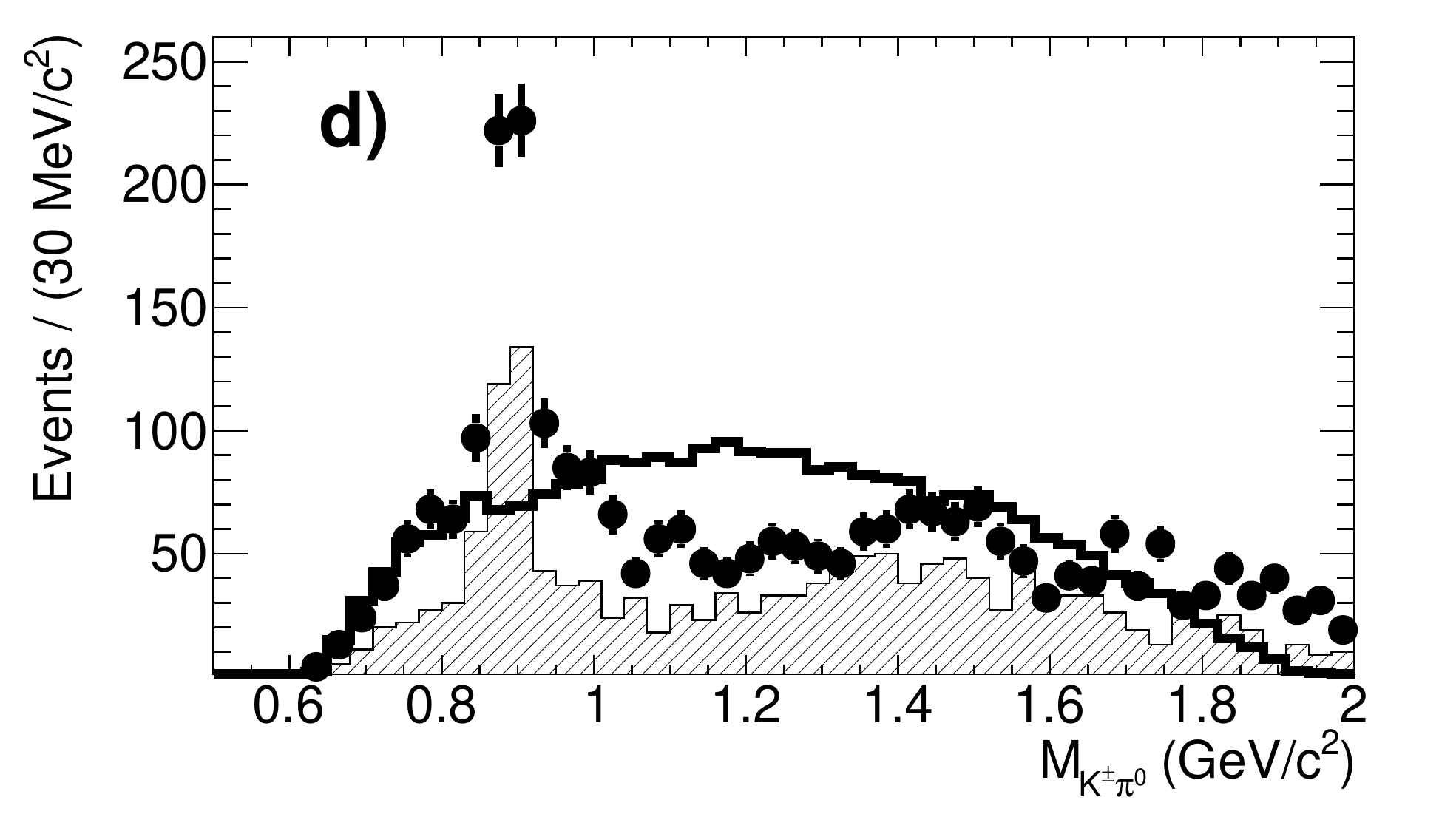}
 \caption{Invariant-mass distributions of $K^{0}_{S}\pi^{0}$ candidates (a and b) and $K^{\pm}\pi^{0}$ candidates (c and d) 
 	  for $\chi_{c0}$ (a and c) and $\chi_{c2}$ (b and d) mass regions, respectively. 
	  Dots represent data, filled histograms represent inclusive MC results, 
	  open histograms show results of signal MC simulations based on a phase-space distribution (arbitrarily scaled).}
 \label{fig:Kpi0}
\end{figure*}

\section{Data analysis}

For the identification and selection of $\psi(3686)\rightarrow \gamma \chi_{c0,2} \rightarrow \gamma \pi^{0}\eta_{c}$ events, 
where $\pi^{0}\rightarrow\gamma\gamma$, $\eta_{c}\rightarrow K^{0}_{S}K^{\pm}\pi^{\mp}$, the $K^{0}_{S}$ is reconstructed in
its decay mode to $\pi^+\pi^-$, resulting in the final state $3\gamma 3\pi K$, where $K$ and $\pi$ are charged.

\subsection{Event selection}
Charged tracks are reconstructed from the MDC hits. For each charged-particle track, its polar angle must satisfy $|\cos\theta| < 0.93$. 
A good charged-particle track (excluding those coming from a $K^{0}_{S}$) is required to be within 1~cm of the $e^{+}e^{-}$ annihilation interaction point (IP), 
transverse to the beam line and within 10~cm of the IP along the beam axis.  
%The IP is determined on a run-by-run basis.
Charged-particle identification (PID) is based on combining the energy loss, $dE/dx$, in the MDC
and TOF information to construct PID chi-squared values $\chi^{2}_{PID} (i)$, 
that are calculated for each charged-particle track for each particle hypothesis $i$ (pion, kaon).

Photons are reconstructed from isolated showers in the EMC.
The showers in the angular range between the barrel ($|\cos\theta| < 0.8$) and end caps ($0.86 < |\cos\theta| < 0.92$) are poorly reconstructed 
and excluded from the analysis. 
Good photon candidates must have a minimum energy of 25~(50)~MeV in the barrel (end cap) regions. 
EMC timing requirements are used to further suppress noise and energy depositions unrelated to the event.

Events with four charged-particle tracks with a net charge of zero and at least three good photon candidates are retained for further analysis.

$K^{0}_{S}$ candidates are reconstructed from secondary vertex fits to all the charged-track pairs in an event (with a pion-mass assumption).
Candidates with an invariant mass within 10~MeV/$c^{2}$ of the $K^{0}_{S}$ nominal mass are considered and 
the combination with the smallest chi-squared of the vertex fit is chosen.
The event is kept for further analysis if the secondary vertex is at least 0.5~cm away from the IP. 
The reconstructed four-momenta of the $\pi^+$ and $\pi^-$, corresponding to the $K^{0}_{S}$ decay, are used as input for the 
subsequent kinematic fit. 
To suppress the $K^{0}_{S}K^{0}_{S}$ background, the remaining charged-particle tracks are required to not form a good $K^{0}_{S}$ candidate.
The $\pi^{0}$ candidates are reconstructed from pairs of photons with the invariant mass 
$M_{\gamma\gamma}$ in the range $0.11 < M_{\gamma\gamma}/{\rm (GeV/}c^{2}) < 0.16$, with the $M_{\gamma\gamma}$ resolution of about 5~MeV/$c^{2}$.

The $3\gamma 3\pi K$ candidates are then subjected to a four-constraint (4C) kinematic fit, 
with the constraints provided by four-momentum conservation. 
The discrimination of charge-conjugate channels ($K^{0}_{S}K^{+}\pi^{-}$ or $K^{0}_{S}K^{-}\pi^{+}$) 
and the selection of the best photon candidate of the $\psi(3686)\rightarrow\gamma \chi_{c0,2}$ transition 
among multiple candidates are achieved by taking the event with the minimum 
$\chi^{2}=\chi^{2}_{4C}+\chi^{2}_{PID}(K)+\chi^{2}_{PID}(\pi)$, 
where $\chi^{2}_{4C}$ is the chi-squared of the 4C kinematic fit.
The $\pi^0$ is reconstructed from the two-photon combination with an invariant mass closest to that of a neutral pion.
Events with $\chi^{2}_{4C} < 50$ and with an invariant mass of the reconstructed $\eta_{c}$, $M_{K^{0}_{S}K^{\pm}\pi^{\mp}}$, in the range 
$2.70 < M_{K^{0}_{S}K^{\pm}\pi^{\mp}}/{\rm(GeV/}c^{2}) < 3.30$ are accepted for further analysis. 
The maximum value of $\chi^{2}_{4C}$ is determined by optimizing the statistical significance $S/\sqrt{S + B}$ in the $\eta_{c}$ signal region, 
where $S~(B)$ is the number of signal (background) events obtained from the signal (inclusive) MC samples.
For the estimate of $S$, the branching fractions of $\chi_{c0,2}\rightarrow \pi^{0}\eta_{c}$ are assumed to be $10^{-3}$ 
in analogy with the isospin-violating process $\psi(3686)\rightarrow\pi^{0}J/\psi$~\cite{pi0jpsi2012}. 
The signal region is defined as $2.90 < M_{K^{0}_{S}K^{\pm}\pi^{\mp}}/{\rm(GeV/}c^{2}) < 3.05$.
The $\chi_{c0}$ and $\chi_{c2}$ signal regions are defined for transition-photon candidates with energies 
in the $\gamma\pi^{0}K^{0}_{S}K^{\pm}\pi^{\mp}$ center-of-mass system, $E_{\gamma}$, in the ranges of 
$0.24 < E_{\gamma}/\rm(GeV) < 0.28$ and $0.10 < E_{\gamma}/\rm(GeV) < 0.15$, respectively. 

\begin{figure*}[ht]
 \centering
  \includegraphics[angle=0,scale=0.33]{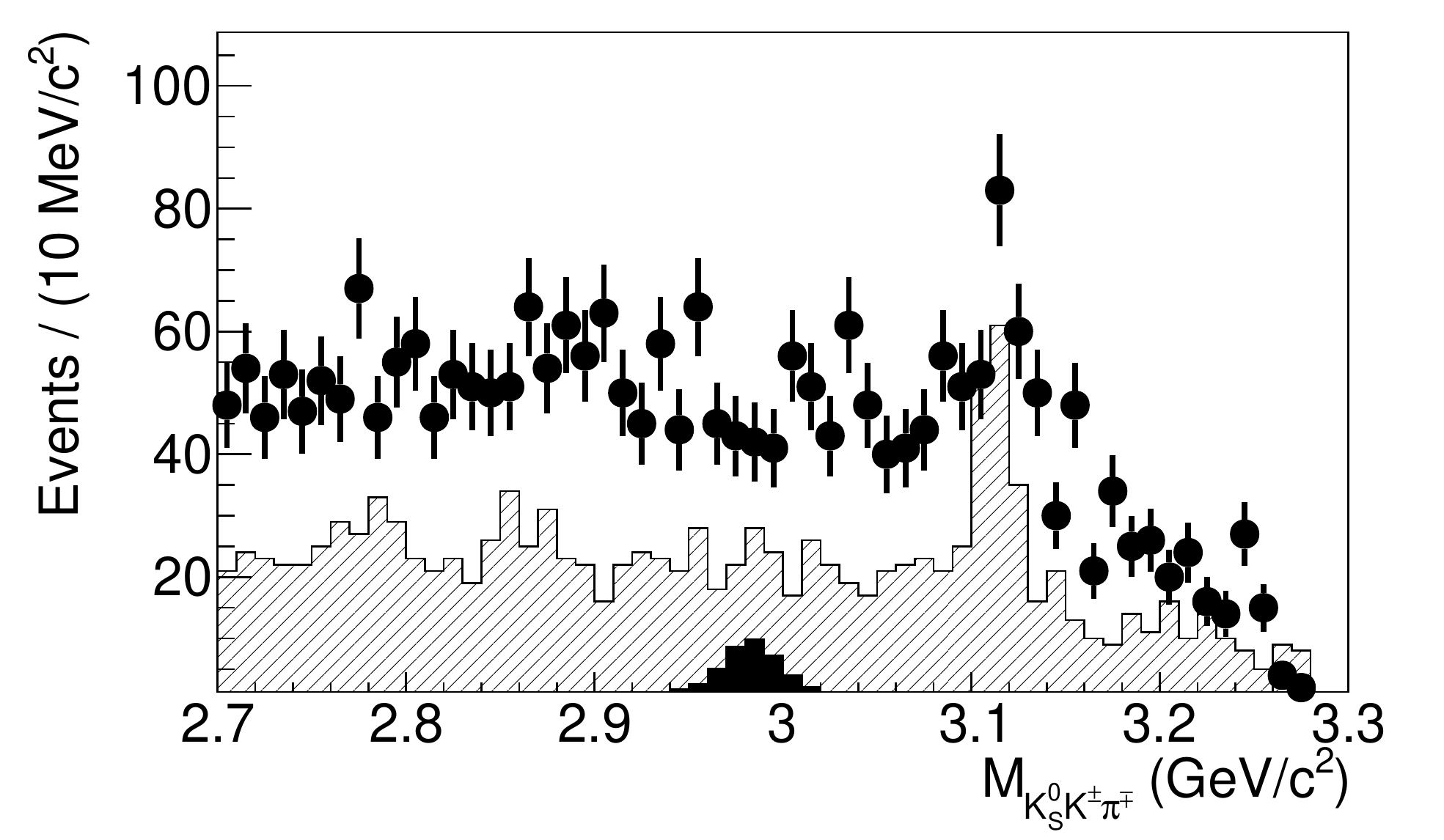}
  \includegraphics[angle=0,scale=0.33]{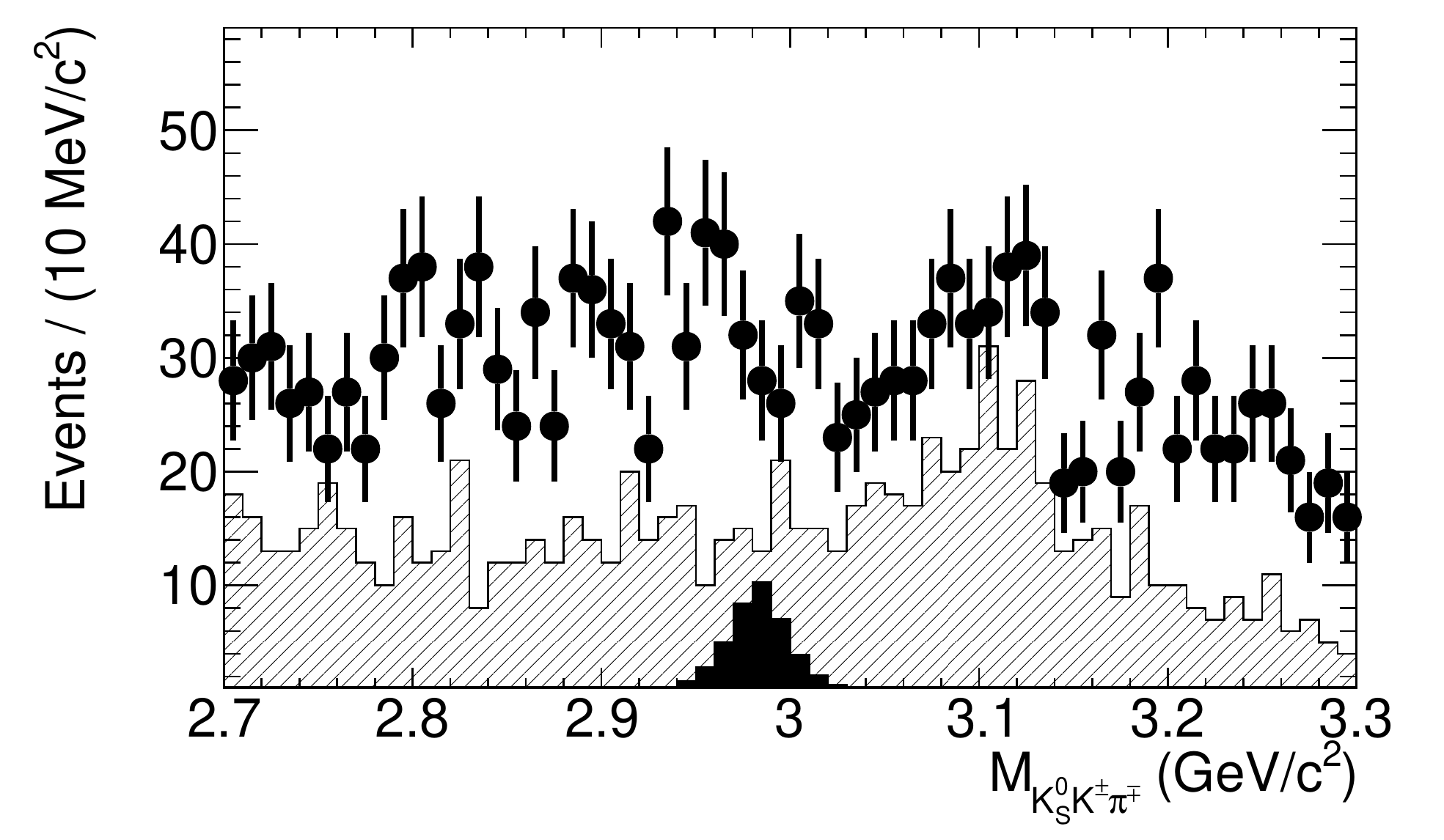}\\
  \caption{Invariant-mass distributions of $K^{0}_{S}K^\pm\pi^\mp$. Left: $\chi_{c0} \rightarrow \pi^{0} \eta_{c}$, 
	right: $\chi_{c2} \rightarrow \pi^{0} \eta_{c}$; $\eta_{c} \rightarrow K^{0}_{S}K^\pm\pi^\mp$. 
	Dots represent data, filled histograms represent the inclusive (light) and arbitrarily scaled signal (dark) MC results. 
	The peak around 3.12~GeV/$c^{2}$ is due to the $\psi(3686)\rightarrow\pi^{0}\pi^{0}J/\psi$ background channel.}
  \label{fig:etac_rec}
\end{figure*}

FIG.~\ref{fig:Kpi0} shows the invariant-mass distributions of $K^{0}_{S}\pi^{0}$ candidates [(a) and (b)] 
and $K^{\pm}\pi^{0}$ candidates [(c) and (d)] with $\pi^{0}K^{0}_{S} K^\pm\pi^\mp$ masses 
within the $\chi_{c0}$ [(a) and (c)] and the $\chi_{c2}$ [(b) and (d)] mass regions. 
The most prominent peak (with the highest intensity and narrowest width) stems from decays involving a $K^{*}(892)$. 
Those are evidently background processes, because the channel of interest, $\chi_{c0,2}\rightarrow \pi^0 \eta_{c}$ with 
$\eta_c \rightarrow K^{0}_{S}K^{\pm}\pi^{\mp}$, cannot involve 
$K^{*}(892)^{0(\pm)} \rightarrow K^{0(\pm)}\pi^{0}$ decays, since the latter does not involve a $\pi^{0}$.
The regions $0.84 < M_{K^{0 (\pm)}_{S}\pi^{0}}/{\rm(GeV/}c^{2}) < 0.95$ % and $0.84 < M_{K^{\pm}\pi^{0}}/{\rm(GeV/}c^{2}) < 0.95$ 
are excluded in the further analysis to suppress the background from $K^{*}(892)^{0}$ and $K^{*}(892)^{\pm}$ decays. 
This condition is optimized to obtain the best statistical significance of the signal.

FIG.~\ref{fig:etac_rec} shows the invariant-mass distributions of $K^{0}_{S}K\pi$ for candidate events 
with $K^{0}_{S} K^\pm\pi^\mp$ masses corresponding to $\chi_{c0,2} \rightarrow \pi^{0}\eta_{c}$ transitions. 
The data show no visible peak in the $\eta_{c}$ signal region. 
In the present analysis, upper limits at the 90$\%$ confidence level (C.L.) for the transitions $\chi_{c0,2} \rightarrow \pi^{0}\eta_{c}$ are determined.  
Inclusive MC results do not reproduce the number of events found in the data, but reproduce the shape of the invariant-mass distributions of $K^{0}_{S}K\pi$ quite well. 
The discrepancies between data and inclusive MC are primarily due to inaccuracies of branching fractions in the generator and due 
to mismatches in their corresponding decay dynamics. 
We note, however, that the inclusive MC data have solely been used to identify background sources 
and to optimize selection criteria. The selection criteria were optimized by assuming a 
branching fraction of 10$^{-3}$ for the $\chi_{cJ}\rightarrow \pi^0 \eta_c$ channels in 
combination with background taken from the inclusive MC sample. We varied 
the signal-to-background ratio by a factor of 2, adjusted our selection criteria accordingly, 
and found a negligible effect on the precision of our final result.
  
%This feature allows us to use the results of inclusive MC simulations to optimize the selection criteria.  

Efficiencies are calculated using the signal MC simulation samples and are found to be 5.8\% and 8.6\% for the $\chi_{c0}$ and $\chi_{c2}$ channels, respectively.

\subsection{Background studies}

Background events from $\psi(3686)$ decays are studied with the inclusive MC sample. These studies showed that the channel 
$\psi(3686) \rightarrow \pi^{0}\pi^{0} J/\psi$, $J/\psi \rightarrow K^{0}_{S}K^{\pm}\pi^{\mp}$ results in a peak 
around 3.12~GeV/$c^{2}$ in the $K^{0}_{S}K\pi$ invariant-mass spectrum as can be observed from Fig.~\ref{fig:etac_rec}. 
In this type of transition, one of the photons originating from $\pi^{0}$ decays may escape, 
which causes a smaller total energy for the event. The kinematic fit increases the energy of the charged-particle tracks, 
which results in a shift in the invariant mass from 
3.10~GeV/$c^{2}$ to 3.12~GeV/$c^{2}$.
% The contribution from these channels is calculated using signal MC simulations and taking the branching fractions from the PDG~\cite{PDG2012}, and is found 
% to be 118$\pm$4 and 33$\pm$2 for the $\chi_{c0}$ and $\chi_{c2}$ channel, respectively.
This decay channel is taken into account in the final fit to the invariant-mass spectrum of $K^{0}_{S}K\pi$ as described below.

The major background contribution stems from the channels $\psi(3686)\rightarrow\gamma\chi_{cJ}$, $\chi_{cJ}\rightarrow\pi^{0}K^{0}_{S}K^{\pm}\pi^{\mp}$.
These channels have final states that are kinematically identical to the signal of interest and, therefore, cannot be removed easily.
Partly, this type of background has been suppressed by vetoing $K^*(892)^0$ signals via a cut on the $K^{0,\pm}_{S}\pi^{0}$ mass 
since the background decay, $\chi_{c0,2} \rightarrow \pi^0 K^0_S K^\pm\pi^\mp$, contains intermediate $K^*(892)$ resonances, 
as discussed earlier. 
We note that the remaining contribution of this type does not result in a peaking background in the signal region.
% Using the exclusive MC simulations and taking the corresponding branching fractions from the PDG~\cite{PDG2012}, 
% the contribution of $\chi_{cJ}\rightarrow\pi^{0}K^{0}_{S}K^{\pm}\pi^{\mp}$ 
% channels in the region $2.70<M_{K^{0}_{S}K\pi}~{\rm(GeV/}c^{2})< 3.20$ is found to be 
% 2216$\pm$437 and 1716$\pm$300 events for the $\chi_{c0}$ and $\chi_{c2}$ selection criteria, respectively, 
% while the total number of data events in the same region is 2477$\pm$50 and 1527.50$\pm$39, respectively. 

The background contribution from $e^{+}e^{-} \rightarrow f\bar{f}$ processes, where $f=e,\mu, d, u, s$, 
is studied using the continuum data taken at $\sqrt{s} = 3.65$~GeV, and it is found to be negligible. 
Using the exclusive MC simulations and taking the corresponding branching fractions from the PDG~\cite{PDG2014}, 
the contribution of $\chi_{cJ}\rightarrow\pi^{0}K^{0}_{S}K^{\pm}\pi^{\mp}$ and $\psi(3686) \rightarrow \pi^{0}\pi^{0} J/\psi$ 
channels in the region $2.70<M_{K^{0}_{S}K\pi}/{\rm(GeV/}c^{2})< 3.30$ is found to be 
2260$\pm$340 and 1668$\pm$260 events for the $\chi_{c0}$ and $\chi_{c2}$ selection criteria, respectively, 
where the errors are mainly due to the uncertainties in the branching fractions.
The total number of data events in the same region is 2477$\pm$50 and 1527$\pm$39, respectively.
These are compatible within the uncertainties. 
No significant peaks are observed in the signal region.  

\subsection{Upper limits for the number of signal events}
\begin{figure*}[ht]
 \centering
  \includegraphics[angle=0,scale=0.33]{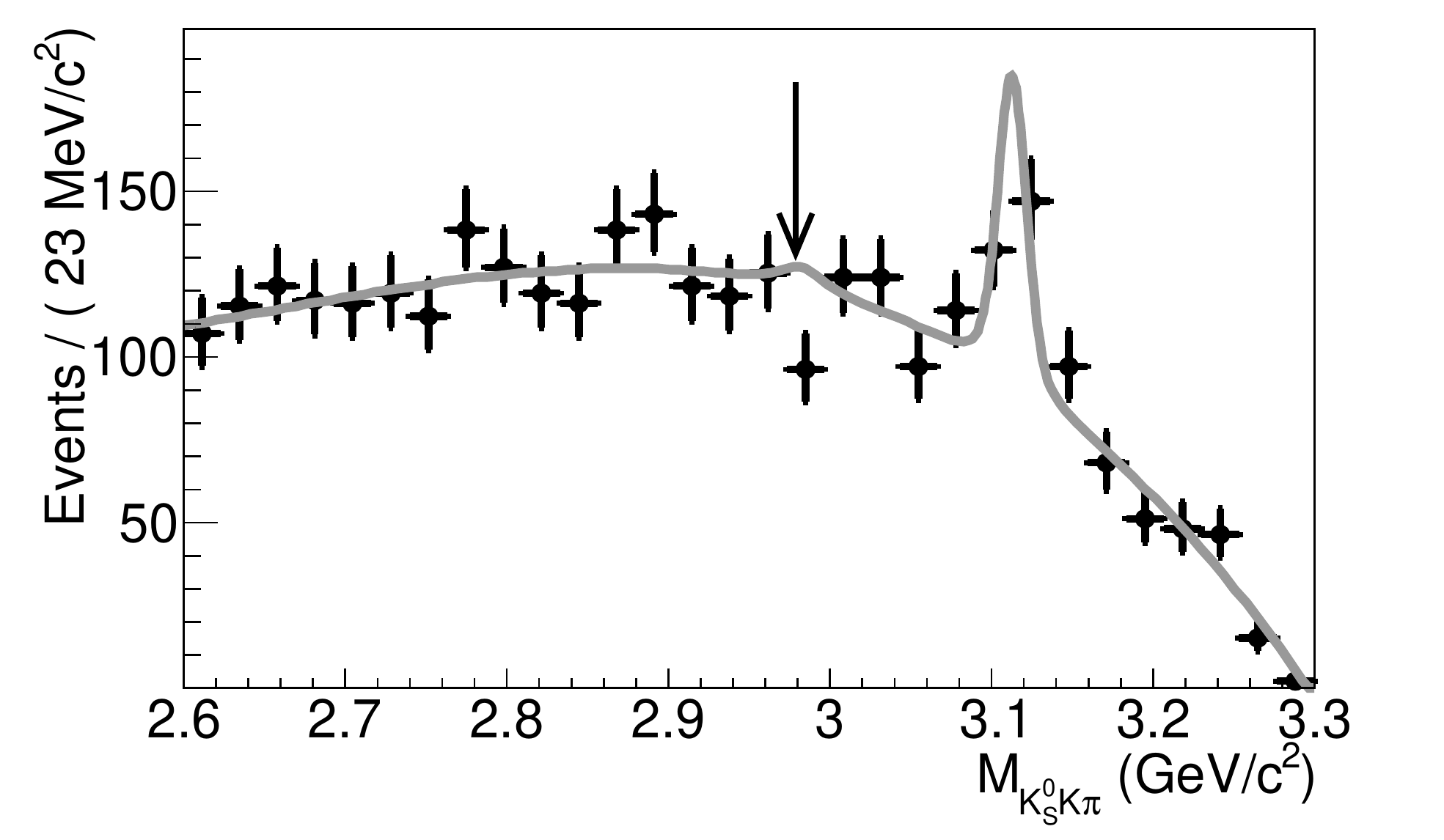}
  \includegraphics[angle=0,scale=0.37]{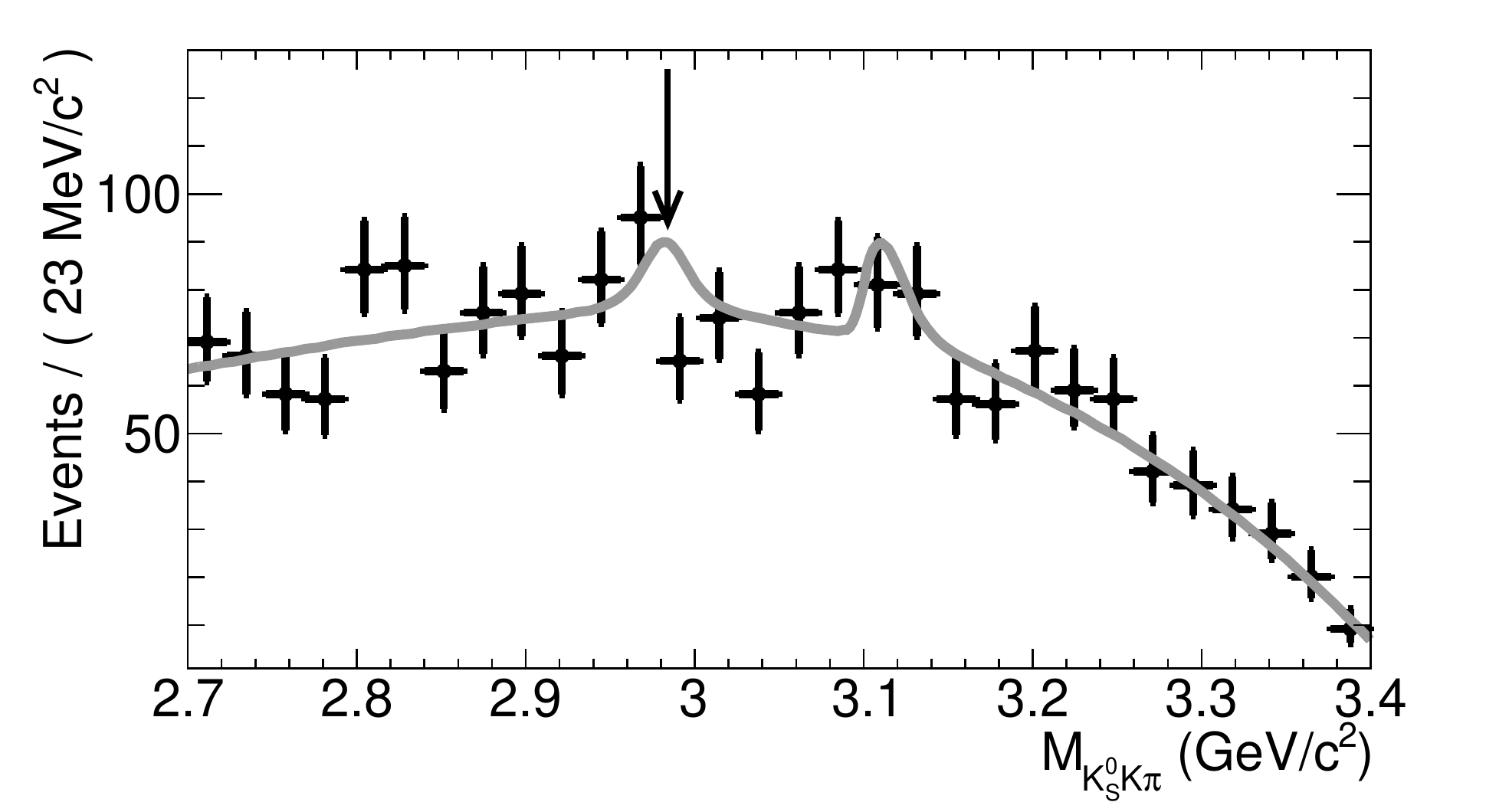}\\
  \caption{Fit to the invariant-mass distributions of $K^{0}_{S}K^\pm\pi^\mp$. Left: $\chi_{c0} \rightarrow \pi^{0} \eta_{c}$, 
	right: $\chi_{c2} \rightarrow \pi^{0} \eta_{c}$; $\eta_{c} \rightarrow K^{0}_{S}K^\pm\pi^\mp$. 
	Dots represent data, lines represent the fit. Arrows indicate the peaks of interest.}
  \label{fig:etac_fit}
\end{figure*}

To extract the number of $\eta_{c}$ events, an unbinned maximum likelihood fit is applied to the 
candidate events with $K^{0}_{S}K^{\pm}\pi^{\mp}$ invariant-mass distributions in the region $2.70 < M_{K_{S}^{0}K\pi}/{\rm(GeV/}c^{2})< 3.30$. 
The $\eta_{c}$ signal is described by a Voigtian function, which is a Breit-Wigner function convoluted with the detector resolution.  
Parameters of the Breit-Wigner function are taken from the PDG~\cite{PDG2014}, 
and the detector resolution is obtained from a fit to the signal MC set. 
These parameters are fixed while fitting the data.
From the background studies, no peaking background is expected in the signal region. 
The smooth background is described by a third-order Chebyshev polynomial. 
A Voigtian function and a Landau plus Gaussian function
are used to describe the structure around 3.12~GeV/$c^{2}$ for the $\chi_{c0}$ and $\chi_{c2}$ mass regions, respectively. 
The line-shape parameters of the structure around 3.12~GeV/$c^{2}$ (for both Voigtian and Landau + Gaussian functions) 
are fixed to the values obtained from the exclusive MC sample.
The MC sample was obtained by simulating the channel 
$\psi(3686)\rightarrow \pi^0\pi^0 J/\psi$ with the exclusive decay $J/\psi\rightarrow K_S^0 K\pi$.    
The total fit results are shown in FIG.~\ref{fig:etac_fit}.
Using the maximum likelihood method, the upper limits on the number of signal events, $N^{UL}$, at the 90\%~C.L. are
found to be 14.1 and 35.9 events for the $\chi_{c0}$ and $\chi_{c2}$ mass regions, respectively.

\section{Systematic uncertainties}

Table~\ref{table:table_etac_syst} summarizes all the systematic uncertainties that are considered in the analysis. 
Below we discuss in more detail the individual sources and the procedure that is used to estimate the errors.

The tracking efficiency for kaons as a function of transverse momentum has been studied using the process 
$J/\psi \to K^{0}_{S}K^{\pm}\pi^{\mp}$, $K^{0}_{S} \to \pi^{+}\pi^{-}$
and the tracking efficiency for pions (not originating from $K^{0}_{S}$) as a function of transverse momentum has been studied using the process 
$\psi(3686) \to \pi^{+}\pi^{-} J/\psi$. 
The difference in efficiencies between data and MC simulations is 2\% for each $K$ or $\pi$ track.
This value is taken as the uncertainty in the tracking efficiency.
The systematic uncertainty due to the $K^{0}_{S}$ reconstruction is 4.0\%, as reported in Ref.~\cite{BAM69}.
The uncertainty in the photon reconstruction is taken as 1\% per photon as reported in Ref.~\cite{PhotonError}. 
In this analysis, there are in total three photons in the final state, which yields a total systematic uncertainty due to the photon 
reconstruction of 3\%.

Some differences are observed for the $\chi^{2}_{4C}$ distributions between data and MC simulations. 
These differences are mainly due to inconsistencies in the charged-track parameters between data and MC simulations. 
We apply correction factors for various $K$ ($\pi$) track parameters that are obtained from the control data samples $J/\psi \to \phi \pi^{+} \pi^{-}$, $\phi \to K^{+}K^{-}$. 
The correction factors are used for smearing the MC simulation output, so that the pull distributions properly describe those 
of the experimental data. Differences between the detection efficiencies obtained using MC simulations with and without these 
corrections are taken as an estimate for the corresponding systematic uncertainties. 
The uncertainties are 1.2\% and 0.8\% for the $\chi_{c0}$ and $\chi_{c2}$ selection conditions, respectively.

A phase-space (PHSP) model used for MC generation of $\eta_{c} \rightarrow K^{0}_{S}K^{\pm}\pi^{\mp}$ events 
does not include possible intermediate resonances between the final-state particles, for example, $K^{*}_{2}(1430)^{0,\pm}$. 
These resonances are observed in the test sample $\psi(3686) \rightarrow \gamma \eta_{c}, \eta_{c} \rightarrow K^{0}_{S}K^{\pm}\pi^{\mp}$. 
%The use of a purely PHSP decay model can introduce systematic errors in the calculation of the reconstruction efficiencies. 
% and calculation of upper limit of signal events. 
To account for $K^{*}$ resonances, additional MC samples of 
$\psi(3686) \rightarrow \gamma \chi_{c0,2}, \chi_{c0,2} \rightarrow \pi^{0} \eta_{c}, \eta_{c} \rightarrow K^{*}_{2}(1430)K, K^{*}_{2}(1430) \rightarrow K\pi$ 
are generated and the reconstruction efficiencies are calculated. 
The difference between efficiencies obtained with two different generator models is taken as a systematic uncertainty. 
For the $\chi_{c0} \rightarrow \pi^{0} \eta_{c}$ and $\chi_{c2} \rightarrow \pi^{0} \eta_{c}$ decays, the corresponding systematic uncertainties 
are 1.6\% and 2.7\%, respectively. 

The selection of exactly four charged-particle tracks before the vertex cuts can introduce 
an additional systematic uncertainty in the efficiency determination due to the presence of fake tracks from misreconstruction.  
This uncertainty is estimated using a sample of $\psi(3686)\rightarrow\pi^{+}\pi^{-} J/\psi$, 
$J/\psi\rightarrow\gamma\eta_{c}$, $\eta_{c}\rightarrow K^{0}_{S}K^{\pm}\pi^{\mp}$ decays, 
where events with at least six charged-particle tracks are accepted. 
The fraction of events with more than six tracks compared to the number of events 
with exactly six charged-particle tracks are obtained for the data and for a corresponding signal MC sample, 
and the difference in the fractions is found to be 1\%. 
This we take as the systematic uncertainty due to the preselection of exactly four charged-particle tracks.

To estimate the systematic uncertainties due to the resolution of the transition photon, 
a smearing of the energy resolution by 1.5~MeV and a shift of 0.5~MeV are introduced on the MC data. 
The smearing gives a minimum $\chi^{2}$ when comparing data and MC line shapes, 
where the dominant contribution to the data originates from the processes $\psi(3686)\rightarrow\gamma\chi_{cJ}$, $\chi_{cJ}\rightarrow\pi^{0}K^{0}_{S}K^{\pm}\pi^{\mp}$. 
The corresponding detection efficiencies of the channels of interest with the standard selection criteria have been calculated. 
The largest difference between the efficiencies with and without smearing for the $\chi_{c0,2}$ is 0.6\%, 
which we quote as the systematic uncertainty due to the transition-photon resolution. 

The systematic uncertainty due to fitting consists of four parts: uncertainties due to the fitting range, 
the $J/\psi$-related background shape, the $\chi_{c0,2}\rightarrow\pi^{0}K^{0}_{S}K\pi$ background shape, and the signal shape.
The upper limits for the $2.70 <M_{K^{0}_{S}K\pi}/{\rm(GeV/}c^{2})< 3.30$ fitting range with a third-order Chebyshev function and fixed $J/\psi$ parameters 
are taken as the nominal upper limit, $N^{UL}$. 
By varying the fitting ranges ($2.60 <M_{K^{0}_{S}K\pi}/{\rm(GeV/}c^{2})< 3.30$ and 
$2.70 <M_{K^{0}_{S}K\pi}/{\rm(GeV/}c^{2})< 3.40$ for the $\chi_{c0}$ and $\chi_{c2}$ mass regions, respectively),  
we obtained a set of upper limits from which we take the maximum difference as a systematic uncertainty. 
The uncertainty is found to be 5.8\% and 15.0\% for the $\chi_{c0}$ and $\chi_{c2}$ mass regions, respectively. 
By changing the parameters for the $J/\psi$ background from fixed to free, but using the nominal fitting range and the nominal order of Chebyshev polynomial, 
the result of the fitting procedure is obtained, and the relative difference with the nominal upper limit is taken as  
as a systematic error due to the line shape uncertainty of the $J/\psi$-related background. 
This error is found to be 7.3\% and 6.8\% for the $\chi_{c0}$ and $\chi_{c2}$ selection conditions, respectively.
By varying the order of the Chebyshev polynomial from the third to second, 
we obtained a set of upper limits from which we take the relative difference as systematic uncertainty
due to the $\chi_{c0,2}\rightarrow\pi^{0}K^{0}_{S}K\pi$ background line shape.
The uncertainty is found to be 0.1\% and 5.0\% for the $\chi_{c0}$ and $\chi_{c2}$ selection conditions, respectively.
By changing the mean of the Voigtian within 1~MeV/$c^{2}$ and the width within 1~MeV, 
thereby taking conservatively into account the uncertainty in the published mass and width of the $\eta_c$~\cite{PDG2014}, 
sets of upper limits are obtained from which we take the maximum difference with the nominal upper limit 
as systematic uncertainty due to the signal line shape. 
This uncertainty is found to be 2.3\% and 2.5\% for the $\chi_{c0}$ and $\chi_{c2}$ selection conditions, respectively. 
The various systematic uncertainties on the fitting range and the line shape for the signal and background are highly correlated 
due to double counting of possible uncertainty contributions. 
The total systematic uncertainty on fitting is estimated by adding the individual systematic uncertainties in quadrature. 
The total fitting uncertainty is 9.6\% and 17.4\% for the $\chi_{c0}$ and $\chi_{c2}$ selection conditions, respectively.

The systematic uncertainty of the number of $\psi(3686)$ events is estimated to be 0.8\% as reported in~\cite{NumPsip}. 
The uncertainty originating from the trigger efficiency is estimated to be 0.15\%~\cite{Trigger}.

All the uncertainties on the branching fractions of the decaying particles of the channels of interest are obtained from the PDG~\cite{PDG2014}
and are taken into account in the systematic errors of our measurements. The corresponding values can be found in Tab.~\ref{table:table_etac_syst}. 
%The uncertainties on the branching fractions of $\psi(3686) \rightarrow \gamma \chi_{c0,2}$ are 
%3.2\% and 3.9\% for the $\chi_{c0}$ and $\chi_{c2}$ channels, respectively. 
%The uncertainties on the branching fractions of $\eta_{c} \rightarrow K^{0}_{S}K\pi$ and $K^{0}_{S} \rightarrow \pi^{+}\pi^{-}$ processes are 6.3\% and 0.1\%, respectively. 
%The uncertainty on the branching fraction of $\pi^{0}\rightarrow\gamma\gamma$ process is less than 0.1\% and, therefore, it is not considered in the analysis. 

Assuming that all the sources are independent, the total systematic uncertainties $\delta_{0,2}$ are obtained by adding the individual uncertainties in quadrature.

\begin{table}[!ht]
\caption{Summary of all considered systematic uncertainties ($\%$). All uncertainties quoted are estimated to be symmetric.}
\begin{center}
\begin{tabular}{ l  c  c }
\hline \hline
Source & $\chi_{c0} \rightarrow \pi^{0} \eta_{c}$ & $\chi_{c2} \rightarrow \pi^{0} \eta_{c}$ \\ \hline   
Tracking of $K$, $\pi$ & 4.0 & 4.0 \\ 
$K^{0}_{S}$ reconstruction & 4.0 & 4.0 \\
Photon reconstruction & 3.0 & 3.0 \\  
Kinematic 4C fitting & 1.2 & 0.8 \\
PHSP generator model & 1.6 & 2.7 \\ 
Four charged-particle tracks & 1.0 & 1.0 \\
$E_{\gamma}$ resolution & 0.6 & 0.6 \\
Fitting & 9.6 & 17.4 \\ 
Number of $\psi(3686)$ & 0.8 & 0.8 \\ 
Trigger & 0.2 & 0.2 \\
B($\psi(3686)\rightarrow \gamma \chi_{cJ}$)~\cite{PDG2014} & 2.7 & 3.4\\
% PDG2014: chic0:9.99+-0.27%, chic2: 9.11+-0.31%; PDG2012: 9.68+-0.31, 8.72+-0.34 
% old number B($\eta_{c} \rightarrow K^{0}_{S}K\pi$)~\cite{PDG2012} & 6.3 & 6.3 \\ 
B($\eta_{c} \rightarrow K^{0}_{S}K\pi$)~\cite{PDG2014} & 6.8 & 6.8 \\ 
B($K^{0}_{S} \rightarrow \pi^{+}\pi^{-}$)~\cite{PDG2014} & 0.1 & 0.1 \\ \hline
Total $\delta_{0,2}$ & 13.8 & 20.2\\
%before 13.8 and 20.2
\hline\hline
\end{tabular}
\label{table:table_etac_syst}
\end{center}
\end{table}

\section{Results and discussion}
The upper limits on the branching fractions of the $\chi_{cJ}\rightarrow\pi^{0}\eta_{c}$~$(J = 0,2)$ transitions are calculated using:

\begin{equation*}
  %\begin{split}
  B(\chi_{cJ} \to \pi^{0} \eta_{c}) < \\ \frac{N^{UL}_{J}}{N_{\psi}\varepsilon_{J} B(\psi(3686) \rightarrow \gamma \chi_{cJ}) B_{int}  (1 - \delta_{J})},
  %\end{split}
\end{equation*}
where 
$N^{UL}_{J}$ are the upper limits on the number of signal events, 
$\delta_{J}$ is the total systematic uncertainty for the channel with $J = 0,2$,
$\varepsilon_{J}$ is the detection efficiency,
$B_{int} = B(\eta_{c} \rightarrow K^{0}_{S}K^{\pm}\pi^{\mp}) 
\cdot B(K^{0}_{S} \rightarrow \pi^{+}\pi^{-}) \cdot B(\pi^{0} \rightarrow \gamma\gamma) = (1.7 \pm 0.3) \cdot 10^{-2}$~\cite{PDG2014},
and $N_{\psi}$ is the number of $\psi(3686)$ events~\cite{NumPsip}. 
Table~\ref{table:table_etac_res} summarizes the final results of the analysis.

\begin{table}[!ht]
\caption{Summary of the final results for the $\chi_{cJ}$ (J = 0,2) decays.}% $N^{UL}_{J}$ are the upper limits on the number of signal events, $\delta_{J}$ are the total systematic errors.}
\begin{center}
\begin{tabular}{ l l l}
\hline \hline
 & $\chi_{c0} \rightarrow \pi^{0} \eta_{c}$ & $\chi_{c2} \rightarrow \pi^{0} \eta_{c}$ \\ \hline   
$N^{UL}_{J}$ & 14.1 & 35.9 \\
$\varepsilon_{J}$ & 5.8\% & 8.6\% \\ %\hline
$\delta_{J}$ & 13.8\% & 20.2\% \\
%$B(\psi(3686) \rightarrow \gamma \chi_{cJ})$~\cite{PDG2012} & (9.7$\pm$0.3)\% & (8.7$\pm$0.3)\% \\
$B(\chi_{cJ} \rightarrow\pi^{0}\eta_{c})(10^{-3})$ & $<$ 1.6 & $<$ 3.2\\
\hline\hline

\end{tabular}
\label{table:table_etac_res}
\end{center}
\end{table}

In this paper, we presented an analysis with the aim to search for 
the hadronic isospin-violating transitions $\chi_{c0,2} \rightarrow \pi^{0} \eta_{c}$ using $106\times 10^{6}$ 
$\psi(3686)$ events collected by BESIII through $\eta_{c} \rightarrow K^{0}_{S}K^{\pm}\pi^{\mp}$ decays. 
No statistically significant signal is observed and upper limits on the branching fractions for the 
processes $\chi_{c0,2} \rightarrow \pi^{0} \eta_{c}$ have been obtained. The results are
$B(\chi_{c0}\rightarrow \pi^0 \eta_c) < 1.6\times 10^{-3}$ and $B(\chi_{c2}\rightarrow \pi^0 \eta_c) < 3.2\times 10^{-3}$.
These are the first upper limits that have been reported so far. 
These limits might help to constrain nonrelativistic field theories and provide insight in the role of charmed-meson loops 
to the various transitions in charmonium and charmonium-like states. 
Further developments in these theories will be necessary to clarify this aspect. 

%The obtained upper limit on the $B(\chi_{c0,2} \rightarrow \pi^{0} \eta_{c})$ does not 
%contradict the result of Ref.~\cite{Voloshin2012}, obtained in the leading-order QCD 
%multipole expansion. \cite{BESIIIpipietac}, \cite{Lu2007}.

The obtained upper limit on $B(\chi_{c0} \rightarrow \pi^{0} \eta_{c})$ does not 
contradict the theoretical estimate reported by Voloshin~\cite{Voloshin2012} 
of order (few)$\times$10$^{-4}$. In this estimate,
the branching fraction has been derived from a leading-order QCD expansion and related to 
the partial width of the decay $\psi(3686)\rightarrow h_c\pi^0$ under the assumption that 
the overlap integrals for the 2S$\rightarrow$1P and 1P$\rightarrow$1S transitions are of similar value. 
In addition, Voloshin~\cite{Voloshin2012} predicts that the branching fractions of the 
hadronic decays $\chi_{c0}\rightarrow \pi^0 \eta_c$ and $\chi_{c1}\rightarrow \pi^+\pi^-\eta_c$ 
are approximately equal. A comparison of our result with that of an upper limit 
measurement of $B(\chi_{c1}\rightarrow\pi^+\pi^-\eta_c < 3.2\times 10^{-3})$ 
by BESIII~\cite{BESIIIpipietac} does not contradict such a prediction. We note, however,
that an earlier theoretical estimate in the framework of a QCD multipole expansion~\cite{Lu2007}
reported a branching fraction for $\chi_{c1}\rightarrow\pi\pi\eta_c$ of (2.22$\pm$1.24)\%, which
contradicts the earlier BESIII measurement~\cite{BESIIIpipietac} and, under the assumption
made by Voloshin~\cite{Voloshin2012}, our result as well.   

The near-future PANDA experiment~\cite{PhysicsPANDA} at the FAIR facility has the potential 
to find evidence or provide tighter constraints for the isospin-forbidden transitions discussed in this paper by directly populating 
 the $\chi_{c0,2}$ states using an intense antiproton beam on a proton target. 
%A detailed balance calculation based on the published $\chi_{c0,2} \rightarrow \bar{p} p$ 
%branching fractions~\cite{PDG2012} and taking into account the foreseen luminosity of $2\times 10^{32}$~cm$^{-2}$s$^{-1}$ 
%gives a production rate of $7.6\times 10^7$ and $1.2 \times 10^8$ $\chi_{c0,2}$ states, respectively, per month at a center-of-mass 
%energy corresponding to the nominal mass of these states.    
%Particularly, for the transition $\chi_{c0} \rightarrow \pi^{0} \eta_{c}$, 
%the NREFT~\cite{Hanhart2011} predicts a non-negligible contribution of intermediate meson loops compared to the tree-level 
%contribution.

\section{Acknowledgments}

The BESIII collaboration thanks the staff of BEPCII and the IHEP computing center for their strong support. 
This work is supported in part by National Key Basic Research Program of China under Contract No. 2015CB856700; 
National Natural Science Foundation of China (NSFC) under Contracts Nos. 11125525, 11235011, 11322544, 11335008, 11425524; 
the Chinese Academy of Sciences (CAS) Large-Scale Scientific Facility Program; Joint Large-Scale Scientific Facility Funds 
of the NSFC and CAS under Contracts Nos. 11179007, U1232201, U1332201; CAS under Contracts Nos. KJCX2-YW-N29, KJCX2-YW-N45; 
100 Talents Program of CAS; INPAC and Shanghai Key Laboratory for Particle Physics and Cosmology; 
German Research Foundation DFG under Contract No. Collaborative Research Center CRC-1044; Istituto Nazionale di Fisica Nucleare, Italy; 
Ministry of Development of Turkey under Contract No. DPT2006K-120470; Russian Foundation for Basic Research under Contract No. 14-07-91152; 
U. S. Department of Energy under Contracts Nos. DE-FG02-04ER41291, DE-FG02-05ER41374, DE-FG02-94ER40823, DESC0010118; 
U.S. National Science Foundation; University of Groningen (RuG) and the Helmholtzzentrum fuer Schwerionenforschung GmbH (GSI), Darmstadt; 
WCU Program of National Research Foundation of Korea under Contract No. R32-2008-000-10155-0

\bibliography{References}

%merlin.mbs apsrev4-1.bst 2010-07-25 4.21a (PWD, AO, DPC) hacked
%Control: key (0)
%Control: author (8) initials jnrlst
%Control: editor formatted (1) identically to author
%Control: production of article title (-1) disabled
%Control: page (0) single
%Control: year (1) truncated
%Control: production of eprint (0) enabled
\begin{thebibliography}{32}%
\makeatletter
\providecommand \@ifxundefined [1]{%
 \@ifx{#1\undefined}
}%
\providecommand \@ifnum [1]{%
 \ifnum #1\expandafter \@firstoftwo
 \else \expandafter \@secondoftwo
 \fi
}%
\providecommand \@ifx [1]{%
 \ifx #1\expandafter \@firstoftwo
 \else \expandafter \@secondoftwo
 \fi
}%
\providecommand \natexlab [1]{#1}%
\providecommand \enquote  [1]{``#1''}%
\providecommand \bibnamefont  [1]{#1}%
\providecommand \bibfnamefont [1]{#1}%
\providecommand \citenamefont [1]{#1}%
\providecommand \href@noop [0]{\@secondoftwo}%
\providecommand \href [0]{\begingroup \@sanitize@url \@href}%
\providecommand \@href[1]{\@@startlink{#1}\@@href}%
\providecommand \@@href[1]{\endgroup#1\@@endlink}%
\providecommand \@sanitize@url [0]{\catcode `\\12\catcode `\$12\catcode
  `\&12\catcode `\#12\catcode `\^12\catcode `\_12\catcode `\%12\relax}%
\providecommand \@@startlink[1]{}%
\providecommand \@@endlink[0]{}%
\providecommand \url  [0]{\begingroup\@sanitize@url \@url }%
\providecommand \@url [1]{\endgroup\@href {#1}{\urlprefix }}%
\providecommand \urlprefix  [0]{URL }%
\providecommand \Eprint [0]{\href }%
\providecommand \doibase [0]{http://dx.doi.org/}%
\providecommand \selectlanguage [0]{\@gobble}%
\providecommand \bibinfo  [0]{\@secondoftwo}%
\providecommand \bibfield  [0]{\@secondoftwo}%
\providecommand \translation [1]{[#1]}%
\providecommand \BibitemOpen [0]{}%
\providecommand \bibitemStop [0]{}%
\providecommand \bibitemNoStop [0]{.\EOS\space}%
\providecommand \EOS [0]{\spacefactor3000\relax}%
\providecommand \BibitemShut  [1]{\csname bibitem#1\endcsname}%
\let\auto@bib@innerbib\@empty
%</preamble>
\bibitem [{\citenamefont {{Ablikim}}\ \emph {et~al.}(2012)\citenamefont
  {{Ablikim}} \emph {et~al.}}]{pi0jpsi2012}%
  \BibitemOpen
  \bibfield  {author} {\bibinfo {author} {\bibfnamefont {M.}~\bibnamefont
  {{Ablikim}}} \emph {et~al.} (\bibinfo {collaboration} {BESIII
  Collaboration}),\ }\href {\doibase 10.1103/PhysRevD.86.092008} {\bibfield
  {journal} {\bibinfo  {journal} {Phys. Rev. D}\ }\textbf {\bibinfo {volume}
  {86}},\ \bibinfo {pages} {092008} (\bibinfo {year} {2012})}\BibitemShut
  {NoStop}%
\bibitem [{\citenamefont {{Olive}}\ \emph {et~al.}(2014)\citenamefont {{Olive}}
  \emph {et~al.}}]{PDG2014}%
  \BibitemOpen
  \bibfield  {author} {\bibinfo {author} {\bibfnamefont {K.~A.}\ \bibnamefont
  {{Olive}}} \emph {et~al.} (\bibinfo {collaboration} {Particle Data Group}),\
  }\href@noop {} {\bibfield  {journal} {\bibinfo  {journal} {Chin. Phys. C}\
  }\textbf {\bibinfo {volume} {38}},\ \bibinfo {pages} {090001} (\bibinfo
  {year} {2014})}\BibitemShut {NoStop}%
\bibitem [{\citenamefont {Choi}\ \emph {et~al.}(2003)\citenamefont {Choi} \emph
  {et~al.}}]{X3872Belle2003isospin}%
  \BibitemOpen
  \bibfield  {author} {\bibinfo {author} {\bibfnamefont {S.-K.}\ \bibnamefont
  {Choi}} \emph {et~al.} (\bibinfo {collaboration} {Belle Collaboration}),\
  }\href {\doibase 10.1103/PhysRevLett.91.262001} {\bibfield  {journal}
  {\bibinfo  {journal} {Phys. Rev. Lett.}\ }\textbf {\bibinfo {volume} {91}},\
  \bibinfo {pages} {262001} (\bibinfo {year} {2003})}\BibitemShut {NoStop}%
\bibitem [{\citenamefont {Choi}\ \emph {et~al.}(2004)\citenamefont {Choi} \emph
  {et~al.}}]{BELLEX3872}%
  \BibitemOpen
  \bibfield  {author} {\bibinfo {author} {\bibfnamefont {S.-K.}\ \bibnamefont
  {Choi}} \emph {et~al.} (\bibinfo {collaboration} {Belle Collaboration}),\
  }\href@noop {} {\bibfield  {journal} {\bibinfo  {journal} {Phys. Rev. D}\
  }\textbf {\bibinfo {volume} {84}},\ \bibinfo {pages} {052004} (\bibinfo
  {year} {2004})}\BibitemShut {NoStop}%
\bibitem [{\citenamefont {Abulencia}\ \emph {et~al.}(2006)\citenamefont
  {Abulencia} \emph {et~al.}}]{CDFX3872}%
  \BibitemOpen
  \bibfield  {author} {\bibinfo {author} {\bibfnamefont {A.}~\bibnamefont
  {Abulencia}} \emph {et~al.} (\bibinfo {collaboration} {CDF Collaboration}),\
  }\href@noop {} {\bibfield  {journal} {\bibinfo  {journal} {Phys. Rev. Lett.}\
  }\textbf {\bibinfo {volume} {96}},\ \bibinfo {pages} {102002} (\bibinfo
  {year} {2006})}\BibitemShut {NoStop}%
\bibitem [{\citenamefont {Chatrchyan}\ \emph {et~al.}(2013)\citenamefont
  {Chatrchyan} \emph {et~al.}}]{CMSJHEP04}%
  \BibitemOpen
  \bibfield  {author} {\bibinfo {author} {\bibfnamefont {S.}~\bibnamefont
  {Chatrchyan}} \emph {et~al.} (\bibinfo {collaboration} {CMS Collaboration}),\
  }\href@noop {} {\bibfield  {journal} {\bibinfo  {journal} {JHEP}\ }\textbf
  {\bibinfo {volume} {04}},\ \bibinfo {pages} {154} (\bibinfo {year}
  {2013})}\BibitemShut {NoStop}%
\bibitem [{\citenamefont {Close}\ and\ \citenamefont
  {Page}(2004)}]{Close2004119}%
  \BibitemOpen
  \bibfield  {author} {\bibinfo {author} {\bibfnamefont {F.~E.}\ \bibnamefont
  {Close}}\ and\ \bibinfo {author} {\bibfnamefont {P.~R.}\ \bibnamefont
  {Page}},\ }\href {\doibase 10.1016/j.physletb.2003.10.032} {\bibfield
  {journal} {\bibinfo  {journal} {Phys. Lett. B}\ }\textbf {\bibinfo {volume}
  {578}},\ \bibinfo {pages} {119} (\bibinfo {year} {2004})}\BibitemShut
  {NoStop}%
\bibitem [{\citenamefont {T{\"o}rnqvist}(2004)}]{Tornqvist2004209}%
  \BibitemOpen
  \bibfield  {author} {\bibinfo {author} {\bibfnamefont {N.~A.}\ \bibnamefont
  {T{\"o}rnqvist}},\ }\href {\doibase 10.1016/j.physletb.2004.03.077}
  {\bibfield  {journal} {\bibinfo  {journal} {Phys. Lett. B}\ }\textbf
  {\bibinfo {volume} {590}},\ \bibinfo {pages} {209} (\bibinfo {year}
  {2004})}\BibitemShut {NoStop}%
\bibitem [{\citenamefont {Voloshin}(2004)}]{Voloshin2004316}%
  \BibitemOpen
  \bibfield  {author} {\bibinfo {author} {\bibfnamefont {M.}~\bibnamefont
  {Voloshin}},\ }\href {\doibase 10.1016/j.physletb.2003.11.014} {\bibfield
  {journal} {\bibinfo  {journal} {Phys. Lett. B}\ }\textbf {\bibinfo {volume}
  {579}},\ \bibinfo {pages} {316} (\bibinfo {year} {2004})}\BibitemShut
  {NoStop}%
\bibitem [{\citenamefont {Swanson}(2004)}]{Swanson2004197}%
  \BibitemOpen
  \bibfield  {author} {\bibinfo {author} {\bibfnamefont {E.~S.}\ \bibnamefont
  {Swanson}},\ }\href {\doibase 10.1016/j.physletb.2004.07.059} {\bibfield
  {journal} {\bibinfo  {journal} {Phys. Lett. B}\ }\textbf {\bibinfo {volume}
  {598}},\ \bibinfo {pages} {197} (\bibinfo {year} {2004})}\BibitemShut
  {NoStop}%
\bibitem [{\citenamefont {Li}\ and\ \citenamefont {Zhu}(2012)}]{Li2012}%
  \BibitemOpen
  \bibfield  {author} {\bibinfo {author} {\bibfnamefont {N.}~\bibnamefont
  {Li}}\ and\ \bibinfo {author} {\bibfnamefont {S.-L.}\ \bibnamefont {Zhu}},\
  }\href {\doibase 10.1103/PhysRevD.86.074022} {\bibfield  {journal} {\bibinfo
  {journal} {Phys. Rev. D}\ }\textbf {\bibinfo {volume} {86}},\ \bibinfo
  {pages} {074022} (\bibinfo {year} {2012})}\BibitemShut {NoStop}%
\bibitem [{\citenamefont {{Shifman}}\ and\ \citenamefont
  {{Ioffe}}(1980)}]{Ioffe1980}%
  \BibitemOpen
  \bibfield  {author} {\bibinfo {author} {\bibfnamefont {M.~A.}\ \bibnamefont
  {{Shifman}}}\ and\ \bibinfo {author} {\bibfnamefont {B.~L.}\ \bibnamefont
  {{Ioffe}}},\ }\href@noop {} {\bibfield  {journal} {\bibinfo  {journal} {Phys.
  Lett. B}\ }\textbf {\bibinfo {volume} {95}},\ \bibinfo {pages} {99} (\bibinfo
  {year} {1980})}\BibitemShut {NoStop}%
\bibitem [{\citenamefont {{Weinberg}}(1977)}]{Weinberg1977}%
  \BibitemOpen
  \bibfield  {author} {\bibinfo {author} {\bibfnamefont {S.}~\bibnamefont
  {{Weinberg}}},\ }\href@noop {} {\bibfield  {journal} {\bibinfo  {journal}
  {Trans. New York Acad. Sci.}\ }\textbf {\bibinfo {volume} {38}},\ \bibinfo
  {pages} {185} (\bibinfo {year} {1977})}\BibitemShut {NoStop}%
\bibitem [{\citenamefont {{Guo}}\ \emph {et~al.}(2009)\citenamefont {{Guo}},
  \citenamefont {{Hanhart}},\ and\ \citenamefont
  {{Mei{\ss}ner}}}]{Hanhart2009}%
  \BibitemOpen
  \bibfield  {author} {\bibinfo {author} {\bibfnamefont {F.-K.}\ \bibnamefont
  {{Guo}}}, \bibinfo {author} {\bibfnamefont {C.}~\bibnamefont {{Hanhart}}}, \
  and\ \bibinfo {author} {\bibfnamefont {U.-G.}\ \bibnamefont
  {{Mei{\ss}ner}}},\ }\href {\doibase 10.1103/PhysRevLett.103.082003}
  {\bibfield  {journal} {\bibinfo  {journal} {\prl}\ }\textbf {\bibinfo
  {volume} {103}},\ \bibinfo {eid} {082003} (\bibinfo {year}
  {2009})}\BibitemShut {NoStop}%
\bibitem [{\citenamefont {Guo}\ \emph {et~al.}(2010)\citenamefont {Guo},
  \citenamefont {Hanhart},\ and\ \citenamefont {Mei{\ss}ner}}]{ErratumHanhart}%
  \BibitemOpen
  \bibfield  {author} {\bibinfo {author} {\bibfnamefont {F.-K.}\ \bibnamefont
  {Guo}}, \bibinfo {author} {\bibfnamefont {C.}~\bibnamefont {Hanhart}}, \ and\
  \bibinfo {author} {\bibfnamefont {U.-G.}\ \bibnamefont {Mei{\ss}ner}},\
  }\href {\doibase 10.1103/PhysRevLett.104.109901} {\bibfield  {journal}
  {\bibinfo  {journal} {Phys. Rev. Lett.}\ }\textbf {\bibinfo {volume} {104}},\
  \bibinfo {pages} {109901(E)} (\bibinfo {year} {2010})}\BibitemShut {NoStop}%
\bibitem [{\citenamefont {{Guo}}\ \emph {et~al.}(2010)\citenamefont {{Guo}},
  \citenamefont {{Hanhart}}, \citenamefont {{Li}}, \citenamefont
  {{Mei{\ss}ner}},\ and\ \citenamefont {{Zhao}}}]{Hanhart2010}%
  \BibitemOpen
  \bibfield  {author} {\bibinfo {author} {\bibfnamefont {F.-K.}\ \bibnamefont
  {{Guo}}}, \bibinfo {author} {\bibfnamefont {C.}~\bibnamefont {{Hanhart}}},
  \bibinfo {author} {\bibfnamefont {G.}~\bibnamefont {{Li}}}, \bibinfo {author}
  {\bibfnamefont {U.-G.}\ \bibnamefont {{Mei{\ss}ner}}}, \ and\ \bibinfo
  {author} {\bibfnamefont {Q.}~\bibnamefont {{Zhao}}},\ }\href {\doibase
  10.1103/PhysRevD.82.034025} {\bibfield  {journal} {\bibinfo  {journal}
  {\prd}\ }\textbf {\bibinfo {volume} {82}},\ \bibinfo {eid} {034025} (\bibinfo
  {year} {2010})}\BibitemShut {NoStop}%
\bibitem [{\citenamefont {{Ablikim}}\ \emph
  {et~al.}(2013{\natexlab{a}})\citenamefont {{Ablikim}} \emph
  {et~al.}}]{NumPsip}%
  \BibitemOpen
  \bibfield  {author} {\bibinfo {author} {\bibfnamefont {M.}~\bibnamefont
  {{Ablikim}}} \emph {et~al.} (\bibinfo {collaboration} {BESIII
  Collaboration}),\ }\href@noop {} {\bibfield  {journal} {\bibinfo  {journal}
  {Chin. Phys. C}\ }\textbf {\bibinfo {volume} {37}} (\bibinfo {year}
  {2013}{\natexlab{a}})}\BibitemShut {NoStop}%
\bibitem [{\citenamefont {{Ablikim}}\ \emph {et~al.}(2010)\citenamefont
  {{Ablikim}} \emph {et~al.}}]{BESIIINIM}%
  \BibitemOpen
  \bibfield  {author} {\bibinfo {author} {\bibfnamefont {M.}~\bibnamefont
  {{Ablikim}}} \emph {et~al.} (\bibinfo {collaboration} {BESIII
  Collaboration}),\ }\href {\doibase 10.1016/j.nima.2009.12.050} {\bibfield
  {journal} {\bibinfo  {journal} {Nuclear Instruments and Methods in Physics
  Research A}\ }\textbf {\bibinfo {volume} {614}},\ \bibinfo {pages} {345}
  (\bibinfo {year} {2010})}\BibitemShut {NoStop}%
\bibitem [{\citenamefont {Deng}\ \emph {et~al.}(2006)\citenamefont {Deng} \emph
  {et~al.}}]{BOOST}%
  \BibitemOpen
  \bibfield  {author} {\bibinfo {author} {\bibfnamefont {Z.~Y.}\ \bibnamefont
  {Deng}} \emph {et~al.},\ }\href@noop {} {\bibfield  {journal} {\bibinfo
  {journal} {Chin. Phys. C}\ }\textbf {\bibinfo {volume} {30}} (\bibinfo {year}
  {2006})}\BibitemShut {NoStop}%
\bibitem [{\citenamefont {{Jadach}}\ \emph {et~al.}(2000)\citenamefont
  {{Jadach}}, \citenamefont {{Ward}},\ and\ \citenamefont {{Was}}}]{KKMC}%
  \BibitemOpen
  \bibfield  {author} {\bibinfo {author} {\bibfnamefont {S.}~\bibnamefont
  {{Jadach}}}, \bibinfo {author} {\bibfnamefont {B.~F.~L.}\ \bibnamefont
  {{Ward}}}, \ and\ \bibinfo {author} {\bibfnamefont {Z.}~\bibnamefont
  {{Was}}},\ }\href {\doibase 10.1016/S0010-4655(00)00048-5} {\bibfield
  {journal} {\bibinfo  {journal} {Computer Physics Communications}\ }\textbf
  {\bibinfo {volume} {130}},\ \bibinfo {pages} {260} (\bibinfo {year}
  {2000})}\BibitemShut {NoStop}%
\bibitem [{\citenamefont {Lange}(2001)}]{EvtGenPaper}%
  \BibitemOpen
  \bibfield  {author} {\bibinfo {author} {\bibfnamefont {D.~J.}\ \bibnamefont
  {Lange}},\ }\href@noop {} {\bibfield  {journal} {\bibinfo  {journal} {Nucl.
  Instrum. Meth. A}\ }\textbf {\bibinfo {volume} {462}},\ \bibinfo {pages}
  {152} (\bibinfo {year} {2001})}\BibitemShut {NoStop}%
\bibitem [{\citenamefont {{Ping}}(2008)}]{BesEvtGen}%
  \BibitemOpen
  \bibfield  {author} {\bibinfo {author} {\bibfnamefont {R.-G.}\ \bibnamefont
  {{Ping}}},\ }\href {\doibase 10.1088/1674-1137/32/8/001} {\bibfield
  {journal} {\bibinfo  {journal} {Chin. Phys. C}\ }\textbf {\bibinfo {volume}
  {32}},\ \bibinfo {pages} {599} (\bibinfo {year} {2008})}\BibitemShut
  {NoStop}%
\bibitem [{\citenamefont {{Chen}}\ \emph {et~al.}(2000)\citenamefont {{Chen}},
  \citenamefont {{Huang}}, \citenamefont {{Qi}}, \citenamefont {{Zhang}},\ and\
  \citenamefont {{Zhu}}}]{Lundcharm}%
  \BibitemOpen
  \bibfield  {author} {\bibinfo {author} {\bibfnamefont {J.~C.}\ \bibnamefont
  {{Chen}}}, \bibinfo {author} {\bibfnamefont {G.~S.}\ \bibnamefont {{Huang}}},
  \bibinfo {author} {\bibfnamefont {X.~R.}\ \bibnamefont {{Qi}}}, \bibinfo
  {author} {\bibfnamefont {D.~H.}\ \bibnamefont {{Zhang}}}, \ and\ \bibinfo
  {author} {\bibfnamefont {Y.~S.}\ \bibnamefont {{Zhu}}},\ }\href {\doibase
  10.1103/PhysRevD.62.034003} {\bibfield  {journal} {\bibinfo  {journal}
  {\prd}\ }\textbf {\bibinfo {volume} {62}},\ \bibinfo {eid} {034003} (\bibinfo
  {year} {2000})}\BibitemShut {NoStop}%
\bibitem [{\citenamefont {Karl}\ \emph {et~al.}(1976)\citenamefont {Karl},
  \citenamefont {Meshkov},\ and\ \citenamefont {Rosner}}]{AngChi1}%
  \BibitemOpen
  \bibfield  {author} {\bibinfo {author} {\bibfnamefont {G.}~\bibnamefont
  {Karl}}, \bibinfo {author} {\bibfnamefont {S.}~\bibnamefont {Meshkov}}, \
  and\ \bibinfo {author} {\bibfnamefont {J.~L.}\ \bibnamefont {Rosner}},\
  }\href {\doibase 10.1103/PhysRevD.13.1203} {\bibfield  {journal} {\bibinfo
  {journal} {Phys. Rev. D}\ }\textbf {\bibinfo {volume} {13}},\ \bibinfo
  {pages} {1203} (\bibinfo {year} {1976})}\BibitemShut {NoStop}%
\bibitem [{\citenamefont {Kabir}\ and\ \citenamefont {Hey}(1976)}]{AngChi2}%
  \BibitemOpen
  \bibfield  {author} {\bibinfo {author} {\bibfnamefont {P.~K.}\ \bibnamefont
  {Kabir}}\ and\ \bibinfo {author} {\bibfnamefont {A.~J.~G.}\ \bibnamefont
  {Hey}},\ }\href {\doibase 10.1103/PhysRevD.13.3161} {\bibfield  {journal}
  {\bibinfo  {journal} {Phys. Rev. D}\ }\textbf {\bibinfo {volume} {13}},\
  \bibinfo {pages} {3161} (\bibinfo {year} {1976})}\BibitemShut {NoStop}%
\bibitem [{\citenamefont {{Ablikim}}\ \emph
  {et~al.}(2013{\natexlab{b}})\citenamefont {{Ablikim}} \emph
  {et~al.}}]{BAM69}%
  \BibitemOpen
  \bibfield  {author} {\bibinfo {author} {\bibfnamefont {M.}~\bibnamefont
  {{Ablikim}}} \emph {et~al.} (\bibinfo {collaboration} {BESIII
  Collaboration}),\ }\href {\doibase 10.1103/PhysRevD.87.052005} {\bibfield
  {journal} {\bibinfo  {journal} {Phys. Rev. D}\ }\textbf {\bibinfo {volume}
  {87}},\ \bibinfo {pages} {052005} (\bibinfo {year}
  {2013}{\natexlab{b}})}\BibitemShut {NoStop}%
\bibitem [{\citenamefont {{Ablikim}}\ \emph {et~al.}(2011)\citenamefont
  {{Ablikim}} \emph {et~al.}}]{PhotonError}%
  \BibitemOpen
  \bibfield  {author} {\bibinfo {author} {\bibfnamefont {M.}~\bibnamefont
  {{Ablikim}}} \emph {et~al.} (\bibinfo {collaboration} {BESIII
  Collaboration}),\ }\href {\doibase 10.1103/PhysRevD.83.112005} {\bibfield
  {journal} {\bibinfo  {journal} {\prd}\ }\textbf {\bibinfo {volume} {83}},\
  \bibinfo {eid} {112005} (\bibinfo {year} {2011})}\BibitemShut {NoStop}%
\bibitem [{\citenamefont {{Berger}}\ \emph {et~al.}(2010)\citenamefont
  {{Berger}} \emph {et~al.}}]{Trigger}%
  \BibitemOpen
  \bibfield  {author} {\bibinfo {author} {\bibfnamefont {N.}~\bibnamefont
  {{Berger}}} \emph {et~al.} (\bibinfo {collaboration} {BESIII
  Collaboration}),\ }\href {\doibase 10.1088/1674-1137/34/12/001} {\bibfield
  {journal} {\bibinfo  {journal} {Chin. Phys. C}\ }\textbf {\bibinfo {volume}
  {34}},\ \bibinfo {pages} {1779} (\bibinfo {year} {2010})}\BibitemShut
  {NoStop}%
\bibitem [{\citenamefont {Voloshin}(2012)}]{Voloshin2012}%
  \BibitemOpen
  \bibfield  {author} {\bibinfo {author} {\bibfnamefont {M.~B.}\ \bibnamefont
  {Voloshin}},\ }\href {\doibase 10.1103/PhysRevD.86.074033} {\bibfield
  {journal} {\bibinfo  {journal} {Phys. Rev. D}\ }\textbf {\bibinfo {volume}
  {86}},\ \bibinfo {pages} {074033} (\bibinfo {year} {2012})}\BibitemShut
  {NoStop}%
\bibitem [{\citenamefont {{Ablikim}}\ \emph
  {et~al.}(2013{\natexlab{c}})\citenamefont {{Ablikim}} \emph
  {et~al.}}]{BESIIIpipietac}%
  \BibitemOpen
  \bibfield  {author} {\bibinfo {author} {\bibfnamefont {M.}~\bibnamefont
  {{Ablikim}}} \emph {et~al.} (\bibinfo {collaboration} {BESIII
  Collaboration}),\ }\href {\doibase 10.1103/PhysRevD.87.012002} {\bibfield
  {journal} {\bibinfo  {journal} {Phys. Rev. D}\ }\textbf {\bibinfo {volume}
  {87}},\ \bibinfo {pages} {012002} (\bibinfo {year}
  {2013}{\natexlab{c}})}\BibitemShut {NoStop}%
\bibitem [{\citenamefont {Lu}\ and\ \citenamefont {Kuang}(2007)}]{Lu2007}%
  \BibitemOpen
  \bibfield  {author} {\bibinfo {author} {\bibfnamefont {Q.}~\bibnamefont
  {Lu}}\ and\ \bibinfo {author} {\bibfnamefont {Y.-P.}\ \bibnamefont {Kuang}},\
  }\href {\doibase 10.1103/PhysRevD.75.054019} {\bibfield  {journal} {\bibinfo
  {journal} {Phys. Rev. D}\ }\textbf {\bibinfo {volume} {75}},\ \bibinfo
  {pages} {054019} (\bibinfo {year} {2007})}\BibitemShut {NoStop}%
\bibitem [{\citenamefont {{Erni}}\ \emph {et~al.}()\citenamefont {{Erni}} \emph
  {et~al.}}]{PhysicsPANDA}%
  \BibitemOpen
  \bibfield  {author} {\bibinfo {author} {\bibfnamefont {W.}~\bibnamefont
  {{Erni}}} \emph {et~al.} (\bibinfo {collaboration} {PANDA Collaboration}),\
  }\href@noop {} {\ }\Eprint {http://arxiv.org/abs/arXiv:0903.3905}
  {arXiv:0903.3905 [hep-ex]} \BibitemShut {NoStop}%
\end{thebibliography}%

\end{document}